\documentclass[final]{svjour3}
\usepackage{rotating}
\usepackage{amssymb}
\usepackage{mathptmx}
\usepackage{hyperref}
\hypersetup{
     colorlinks   = true,
     allcolors    =blue
}
\usepackage{graphicx}
\usepackage{caption}
\usepackage{tabularx}
\usepackage[numbers, square, sort]{natbib}

\makeatletter
\journalname{Journal of Low Temperature Physics}
\usepackage{etoolbox}

\begin{document}

\newcommand{\hdblarrow}{H\makebox[0.9ex][l]{$\downdownarrows$}-}
\title{CCAT-prime: Characterization of the First 280 GHz MKID Array for Prime-Cam}

\author{S.K. Choi \textsuperscript{1,2} \and 
C.J. Duell \textsuperscript{1} \and 
J. Austermann \textsuperscript{3}  \and  
%S.C. Chapman \textsuperscript{4}  \and   
N.F. Cothard \textsuperscript{4}  \and  
J. Gao \textsuperscript{3}  \and  
R.G. Freundt \textsuperscript{2}  \and  
C. Groppi \textsuperscript{5} \and
T. Herter \textsuperscript{2} \and 
J. Hubmayr \textsuperscript{3} \and 
Z.B. Huber \textsuperscript{1} \and 
B. Keller \textsuperscript{1} \and 
Y. Li \textsuperscript{1,6} \and
P. Mauskopf \textsuperscript{5} \and
M.D. Niemack \textsuperscript{1,2,6} \and 
T. Nikola \textsuperscript{7} \and 
K. Rossi \textsuperscript{7} \and 
A. Sinclair \textsuperscript{5,8} \and 
G.J. Stacey \textsuperscript{2} \and 
E.M. Vavagiakis \textsuperscript{1} \and
M. Vissers \textsuperscript{3} \and
C. Tucker \textsuperscript{9} \and
E. Weeks \textsuperscript{5} \and
J. Wheeler \textsuperscript{3}
}

\institute{\textsuperscript{1}Department of Physics, Cornell University, Ithaca, NY 14853, USA \\
\textsuperscript{2}Department of Astronomy, Cornell University, Ithaca, NY 14853, USA \\
\textsuperscript{3}Quantum Sensors Group, NIST, Boulder, CO 80305, USA \\
\textsuperscript{4}NASA Goddard Space Flight Center, Greenbelt, MD 20771, USA \\
\textsuperscript{5}School of Earth and Space Exploration, Arizona State University, Tempe, AZ 85287, USA \\
\textsuperscript{6}Kavli Institute at Cornell for Nanoscale Science, Cornell University, Ithaca, NY 14853, USA \\
\textsuperscript{7}Cornell Center for Astrophysics and Planetary Science, Cornell University, Ithaca, NY 14853, USA \\
\textsuperscript{8}Department of Physics and Atmospheric Science, Dalhousie University, Halifax, NS, B3H 4R2, Canada \\
\textsuperscript{9}School of Physics and Astronomy, Cardiff University, CF24 3AA Cardiff, UK \\
%\textsuperscript{4}Argelander Institute for Astronomy, University of Bonn, D-53121 Bonn, Germany \\
%\textsuperscript{5}Kavli Institute for Particle Astrophysics and Cosmology \& Physics Department, Stanford University, Stanford, CA 94305, USA \\
%\textsuperscript{6}Department of Physics, Cornell University, Ithaca, NY 14853, USA \\
%\textsuperscript{7}Cornell Center for Astrophysics and Planetary Science, Cornell University, Ithaca, NY 14853, USA \\
%\textsuperscript{3}Department of Physics, University of Michigan, Ann Arbor, MI 48103, USA \\
%\textsuperscript{5}NASA Goddard Space Flight Center, Greenbelt, MD 20771, USA \\
%\textsuperscript{6}Department of Physics, University of Pennsylvania, Philadelphia, PA 19104, USA  \\
\email{skc98@cornell.edu}
}

\authorrunning
\titlerunning
\maketitle

\begin{abstract}

%FYST, CCAT-prime, 280 GHz MKID array, array data, future plans,

The Prime-Cam receiver on the Fred Young Submillimeter Telescope for the CCAT-prime project aims to address important astrophysical and cosmological questions with sensitive broadband, polarimetric, and spectroscopic measurements. The primary frequency bands in development include 280, 350, and 850 GHz for the polarization-sensitive broadband channels and 210--420 GHz for the spectrometers. Microwave kinetic inductance detectors (MKIDs) are a natural choice of detector technology for the simplicity in multiplexed readout needed for large format arrays at these high frequencies. We present here the initial lab characterization of the feedhorn-coupled 280 GHz polarimetric MKID array, and outline the plans for the subsequent MKID arrays and the development of the testbed to characterize them.


\end{abstract}

\section{Introduction}
The Fred Young Submillimeter Telescope (FYST) for the CCAT-prime project is a new 6-m crossed-Dragone telescope to be located at a 5,600-m altitude site on Cerro Chajnantor in the Atacama Desert. One of the FYST receivers is the 1.8-m diameter Prime-Cam, which will address many important scientific questions ranging from star formation to Big Bang cosmology with sensitive broadband, polarimetric, and spectroscopic measurements in the millimeter to submillimeter regime \cite{ccat/2021}. 

Low temperature detectors continue to play pivotal roles in advancing astronomical measurements. Transition edge sensor (TES) bolometers have been a crucial detector technology enabling sensitive ground based observations in the millimeter regime, leading to many improved cosmological constraints \cite{bicep/2021, dutcher/2021, choi/2020b}. Some upcoming millimeter observations will be made using TES arrays with SQUID-coupled frequency domain multiplexed (FDM) readout \cite{SO/2019}, the complexity of which typically limits the number of detectors and hence the sensitivity at higher frequencies.

Smaller pixels lead to worse individual detector sensitivity (due to lower spillover efficiency) but allow for packing in more detectors for a better total instrument sensitivity \cite{hill/2018}. Optimizing detector size and count to maximize the total instrument sensitivity in the millimeter to submillimeter regime leads to detector density that are much higher than even the most advanced FDM readout promises to handle for TES arrays in the near future \cite{hill/2018, choi/2020a, mccarrick/2021}. 

Microwave kinetic inductance detectors (MKIDs) are a detector technology that is rapidly maturing for astronomical observations \cite{hubmayr/2015,austermann/2018}. The natural FDM scheme for MKIDs simplifies the readout architecture tremendously compared to that of the TES arrays. This makes MKIDs a better detector technology for measurements above 200 GHz with Prime-Cam \cite{choi/2020a}. 

We describe the first light 280 GHz MKID array for Prime-Cam and present its initial laboratory characterization. In particular, we show a measurement that is an important first step of an array post-processing method for improving yield, and the effect of magnetic shielding for the array. We also outline the plans for the subsequent MKID arrays for Prime-Cam and the development of the testbed to characterize them.

%Introduction on FYST, survey, Prime-Cam, Mod-Cam, instrument modules, MKID arrays, readout plans, receiver plans

\begin{figure}[t]
    \centering
%    \begin{subfigure}[b]{0.99\textwidth}
    \includegraphics[width=0.495\textwidth]{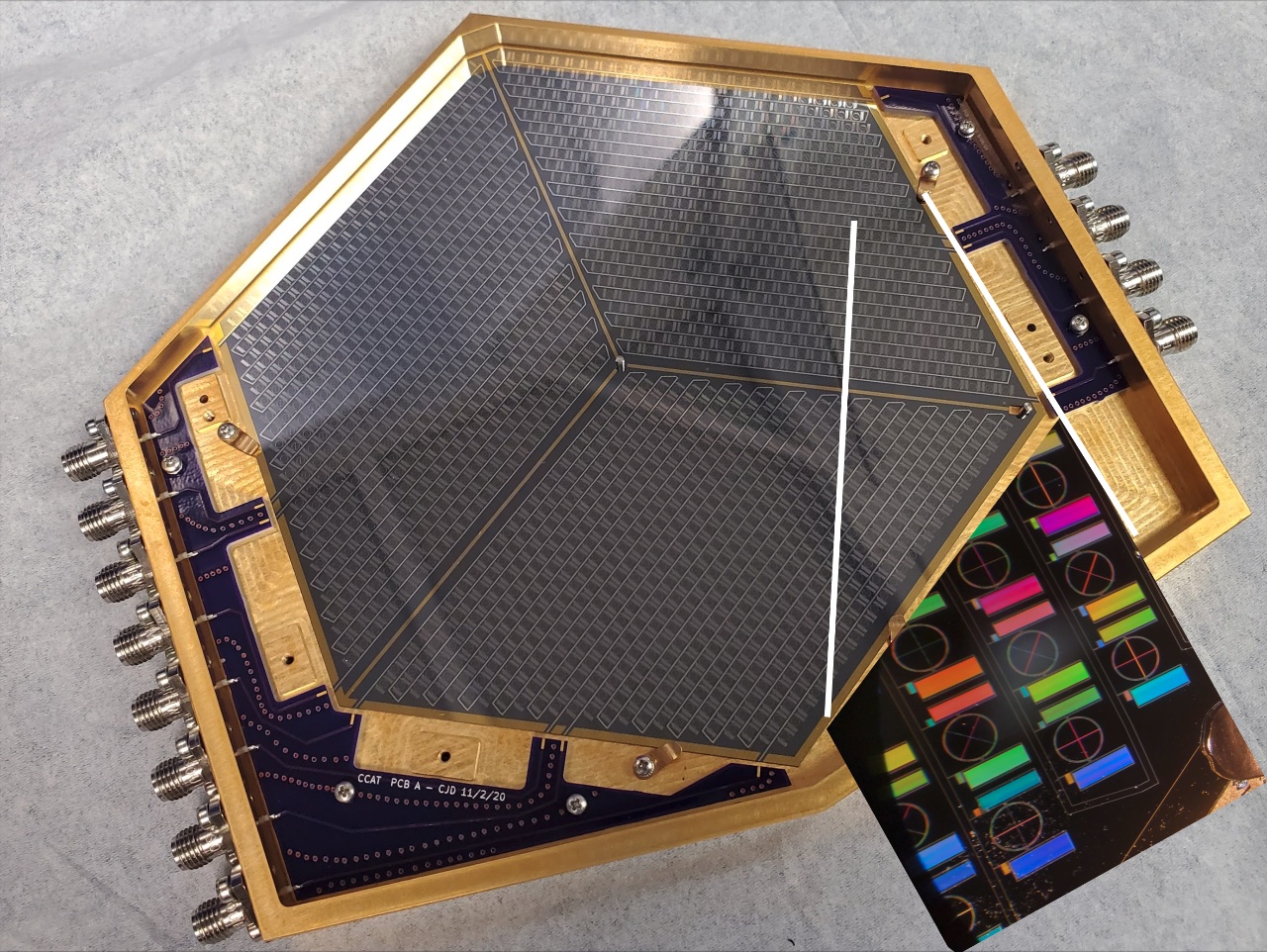}
    \includegraphics[width=0.495\textwidth]{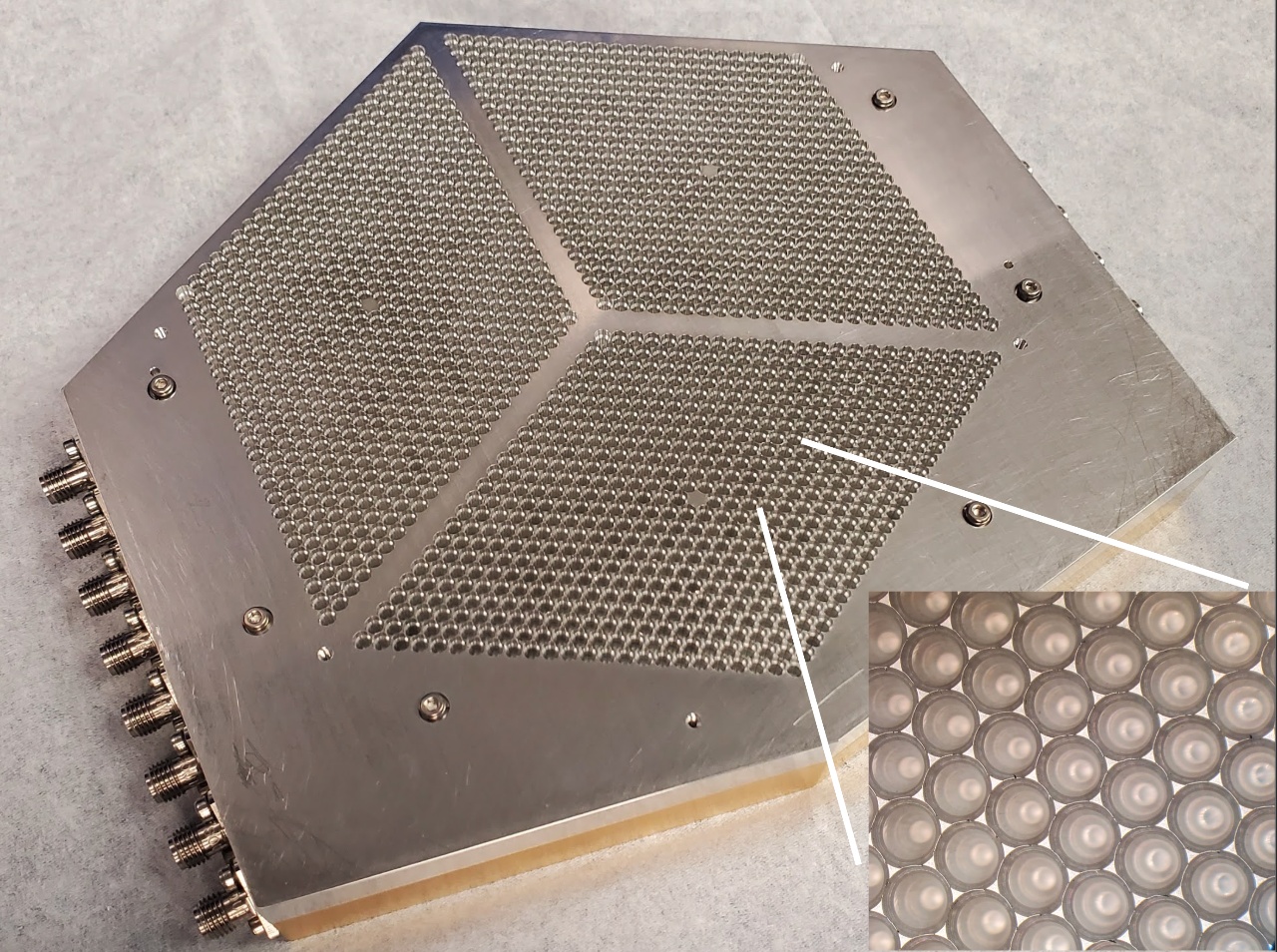}
%\vspace{-0.25in} 
    \caption{Left panel shows the 280 GHz MKID array assembled onto the gold-plated aluminum base. Right panel shows the machined spline-profiled feedhorn array mated to the base with alignment pins.}
\label{fig:array}
\vspace{-0.18in} 
\end{figure}

\section{Instrument and Test Setup}
The mechanical design for the 280 GHz polarimetric MKID array package is described in \cite{duell/2020}. The array is fabricated on a 150 mm diameter wafer at National Institute of Standards and Technology (NIST Boulder), and uses pixel design based on TolTEC \cite{austermann/2018} but laid out in hexagonal close packing in three rhombuses for a total of 3,456 detectors over six RF feedlines (Figure~\ref{fig:array}). The superconducting material is TiN/Ti/TiN trilayer with a critical temperature of $\sim$1.1 K, used as both the inductor of the resonator and absorber that couples to light from the waveguides and feedhorns. The spline-profiled feedhorns with chokes are machined on aluminum with the Kern EVO CNC milling machine at ASU. %The array is assembled onto the gold-plated aluminum base, aligned with 1.5 mm diameter dowel pins, held down with four BeCu tabs, and aluminum-wirebonded to PCBs. The base shown in the left panel of Figure~\ref{fig:array} holds the array with two alignment pins with a slot on the array for the outer pin to take into account differential thermal contraction. 
The array is assembled onto the gold-plated aluminum base (Figure~\ref{fig:array} left panel), aligned with two 1.5 mm diameter dowel pins, held down with four BeCu tabs, and aluminum-wirebonded to printed circuit boards (PCBs). To compensate for differential thermal contraction, the outer dowel pin hole on the array is elongated.
There are 18 pogo pins placed on the three gaps separating the three rhombuses on the aluminum feedhorns to softly press the array against the base to prevent microphonics. The feedhorns are placed on the base with two other alignment pins, thereby aligning both the array and the feehorns to the base to $\pm15$ $\mu$m. The flatness across the feedhorn array is $\sim$13 $\mu$m. Precision stand-offs set the space between the array and the chokes to be $\sim$110 $\mu$m when cooled to 100 mK. Prior to assembly and testing with the MKID array, mechanical and cryogenic validations were done using a mechanical test array with the base and the feedhorns.

The array is cooled to a base temperature of 100 mK in a Bluefors dilution refrigerator (SD250), and read out with the second generation Reconfigurable Open Architecture Computing Hardware (ROACH-2) board and intermediate frequency (IF) electronics and firmware from Arizona State University (ASU) \cite{gordon/2016}. The six RF feedlines are tested one at a time using a pair of cold and warm readout lines with an RF switch system at 100 mK. Passive components in the RF chain include attenuators, DC blocks, and high-pass filters. The signal is amplified both at 4 K, by a low noise amplifier from ASU, and at room temperature, by a Mini-Circuits ZX60-43-S+. The low noise amplifier provides 30 dB of gain (with a noise temperature of ~5 K) and the room temperature amplifier provides an additional 20 dB. The IF electronics include input and output variable attenuators that are adjusted to maximize the signal-to-noise ratio while maintaining linearity.

\begin{figure}[t]
    \centering
%    \begin{subfigure}[b]{0.99\textwidth}
    \includegraphics[width=0.99\textwidth]{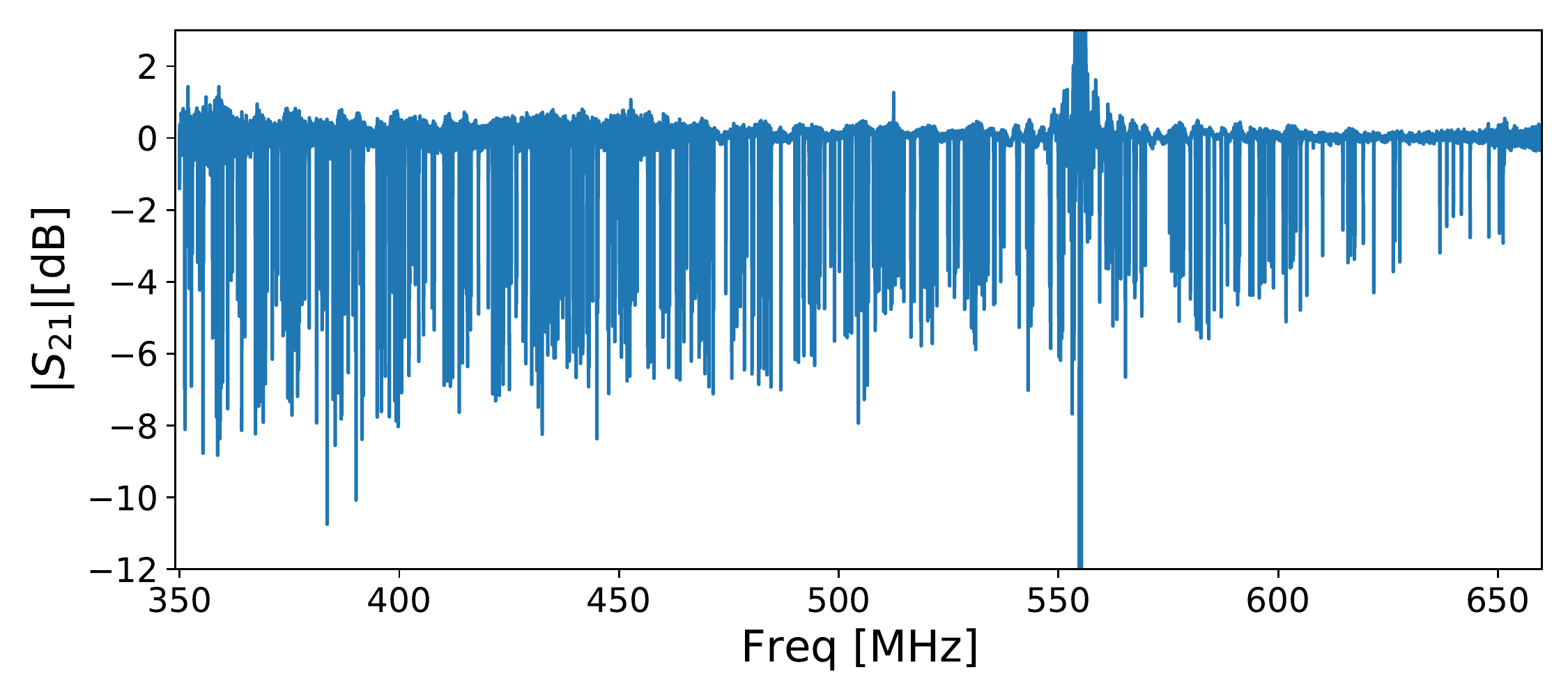}
%\vspace{-0.25in} 
    \caption{The magnitude of the complex forward transmission $|S_{21}|$ at $\sim$105 mK is shown for one feedline. A smooth baseline was removed. An artifact from subtracting the baseline is visible at $\sim$560 MHz, the LO frequency of the IF electronics.}
\label{fig:S21}
\vspace{-0.18in} 
\end{figure}

Various measurement and testbed tools are under active development. They include LED mappers (with components and the design from NIST) to measure the physical position mapping of the resonators, a broad frequency range cold load blackbody source for characterizing the detector responsivity, an IR source with polarizer and XY-stage for measuring beam and polarization efficiency, and a Fourier transform spectrometer for measuring the detector passband.
%In development are LED mappers (from NIST) for physical position mapping of resonators, broad frequency range cold load blackbody source for characterizing detector responsivity, IR source with polarizer and XY-stage for measuring beam and polarization efficiency, and Fourier transform spectrometer for measuring the detector passband. 

While fabrications of MKID arrays continue to improve, the resonator frequencies often deviate from the target frequencies due to imperfections, parasitic reactance, stray magnetic fields, etc. This leads to frequency collisions and degraded yield on MKID arrays. The LED mapper is a PCB with LEDs positioned to control illumination directly over the feedhorns, thereby allowing mapping between the physical position of the detectors and the frequencies of the resonators that respond to the LEDs. This mapping can then be used to fine-tune the resonator frequencies by lithographically trimming the interdigitated capacitors. This can reduce frequency collisions and hence the yield of the array, as demonstrated on a 127-resonator array at NIST \cite{liu/2017}. However, the LED trimming method has yet to be demonstrated on a large format MKID array with feedhorn coupling and choke structures that extend over the capacitors, which could potentially cause fringing fields affecting the resonators. Here we study the sensitivity of the resonators to the precise positioning of the feedhorns with chokes, by measuring the resonator frequency shifts due to removing and replacing the feedhorns over the array. 

% the potential resonator frequency shifts due to
%280 GHz MKID array introduction, feedhorn description including profile (Simon and Henke work), BF cryostat and ROACH-2, ASU IF slice description, readout chain description, RFSoC future, possibly: photos of feedhorns, mag shield, cold load, LED mapper, beam mapper, FTS in development

\begin{figure}[t]
    \centering
%    \begin{subfigure}[b]{0.99\textwidth}
    \includegraphics[width=0.9\textwidth]{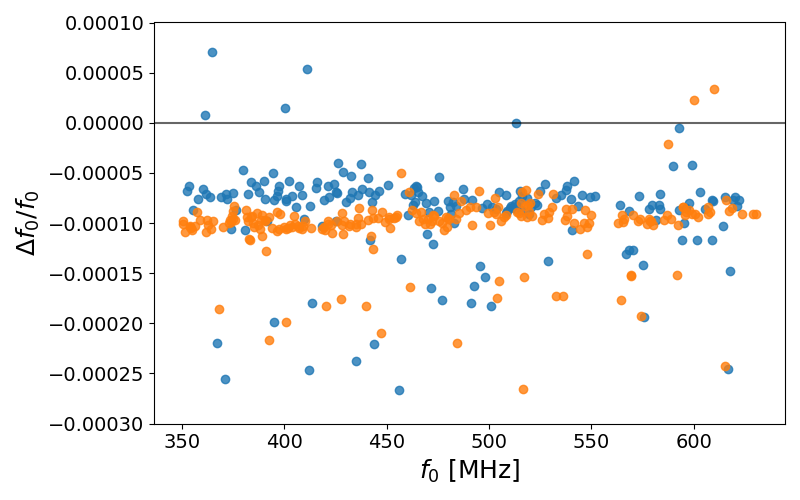}
%\vspace{-0.25in} 
    \caption{Fractional frequency shifts, $\Delta f_0/f_0$, are shown for two feedlines (orange and blue) before and after feedhorns were removed and replaced. The mean and the RMS for the channel shown in blue (orange) is $-9\pm8\times10^{-5}$ ($-10\pm3\times10^{-5}$), after cutting on residuals to parameter fits to remove $\sim$15$\%$ of the detectors. The fractional $f_0$ scatter due to re-assembly procedures shown here is 5--10$\times$ smaller than that from the LED mapper trimming process, $3.5\times10^{-4}$ \cite{liu/2017}.}
\label{fig:delta_f0}
\vspace{-0.18in} 
\end{figure}

\section{Method, Data, and Results}
The primary data we use are the complex forward transmission through the feedline with resonators ($S_{21}$), taken with the ROACH-2 over one of the six RF feedlines at a time:
$$S_{21}(f) = 1 - \frac{Q/Q_c}{1+2jQ\frac{f-f_0}{f_0}},$$
where $f_0$ is the resonant frequency, $Q$ is the measured resonator quality factor, and $Q_c$ is the coupling quality factor for coupling to the RF feedline. The resonator quality factor $Q$ includes $Q_c$, the internal quality factor $Q_i$, and any external loss quality factor $Q_{\rm{loss}}$:
$$\frac{1}{Q} = \frac{1}{Q_c}+\frac{1}{Q_i}+\frac{1}{Q_{\rm{loss}}}.$$

The magnitude of the complex forward transmission $|S_{21}|$ at $\sim$105 mK for one feedline is shown in Figure~\ref{fig:S21}. The resonators trace circles in the complex plane, to which we can fit the resonator parameters in addition to fitting them in the magnitude space. 
%as shown in Figure~\ref{fig:fit_example}. 

%
%\begin{figure}[t]
%    \centering
%%    \begin{subfigure}[b]{0.99\textwidth}
%    \includegraphics[width=0.99\textwidth]{figs/ch4_NewCuBase_107mK_in18out18_amp12mA_shield_recycle_fits_75.pdf}
%%\vspace{-0.25in} 
%    \caption{resonator fit example. find a lower noise one perhaps.}
%\label{fig:fit_example}
%%\vspace{-0.18in} 
%\end{figure}

\begin{figure}[t]
    \centering
%    \begin{subfigure}[b]{0.99\textwidth}
    \includegraphics[height=0.51\textwidth]{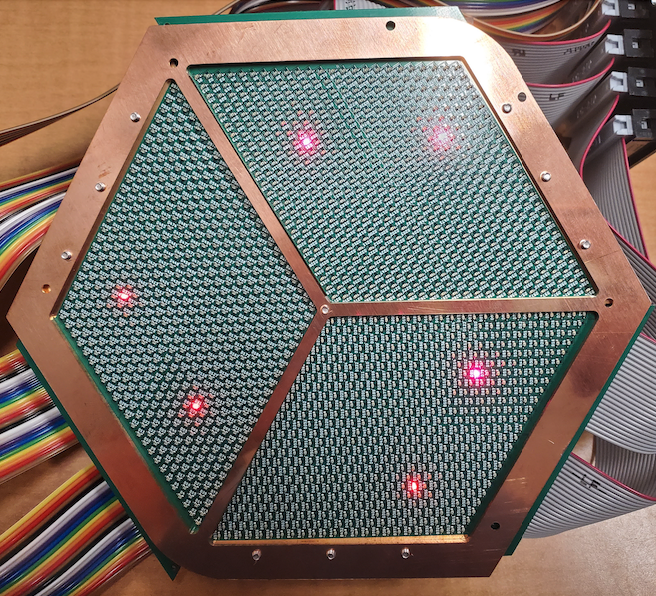}
    \includegraphics[height=0.51\textwidth]{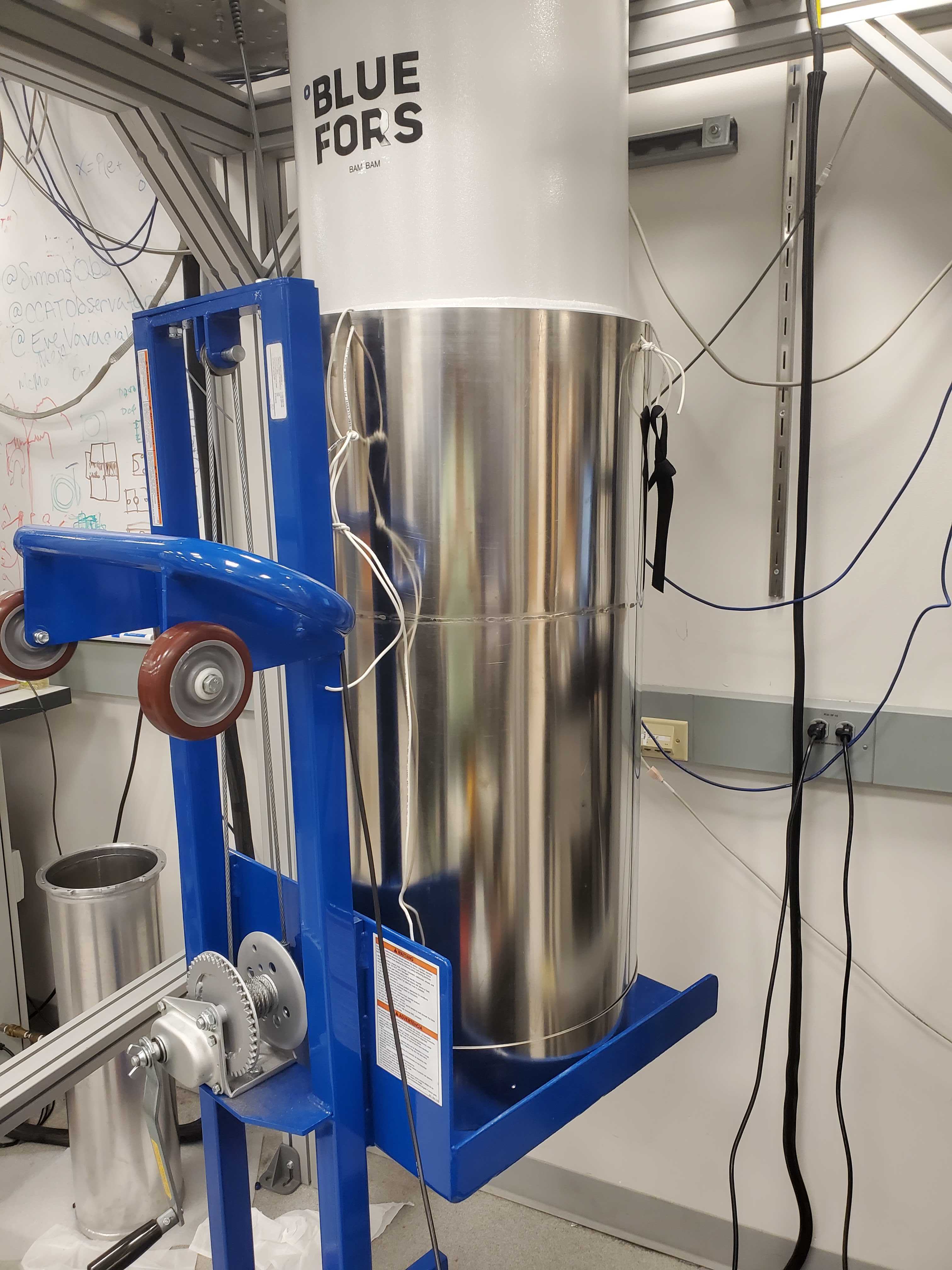}
%\vspace{-0.25in} 
    \caption{Left panel shows the LED mapper array PCBs that will be used to map resonator frequencies to the physical positions. Six LEDs are turned on simultaneously to illuminate one pixel on each of the six feedlines at a time. Right panel shows the external magnetic shield setup used for the shielding test.}
\label{fig:led_shield}
\vspace{-0.18in} 
\end{figure}

We present here the results from the $S_{21}$ data taken before and after removing and replacing the feedhorns over the array, and quantify the level of the shift in resonator frequencies, $\Delta f_0 = f_0^{\rm{after}} - f_0^{\rm{before}}$. The feedhorns are aligned to the base that holds the array with two pins with a precision better than 30 $\mu$m. We test the level of frequency shift that can occur due to this precision. The LED mapper trimming procedure requires a post-fabrication process on the array, and hence it is crucial to maintain the repeatability of resonator frequencies, even after re-assembly of the array with the package, which may result in alignment shifts. We measure $\Delta f_0/f_0$ over two feedlines and find the means and the standard deviations to be $-9\pm8\times10^{-5}$ and $-10\pm3\times10^{-5}$, as shown in Figure~\ref{fig:delta_f0}. The mean shift is partly due to the two datasets being taken at different temperatures, and the scatter is the important quantity. The scatter in fractional $f_0$ spacing can be improved from $2\times10^{-3}$ to $3.5\times10^{-4}$ from the LED mapper trimming process \cite{liu/2017}. Hence the fractional $f_0$ scatter due to re-assembly procedures is 5--10$\times$ smaller than that from the LED mapper trimming process. We repeat this test of removing and replacing the feedhorns one more time and  find good consistency again after the re-assembly. We conclude that the $f_0$ shifts due to re-assembly of the array package are negligible. The LED mappers for the 280 GHz MKID array feedhorns are built (left panel of Figure~\ref{fig:led_shield}), and we will soon proceed with the measurements and corrections.

\begin{figure}[t]
    \centering
%    \begin{subfigure}[b]{0.99\textwidth}
    \includegraphics[width=0.49\textwidth]{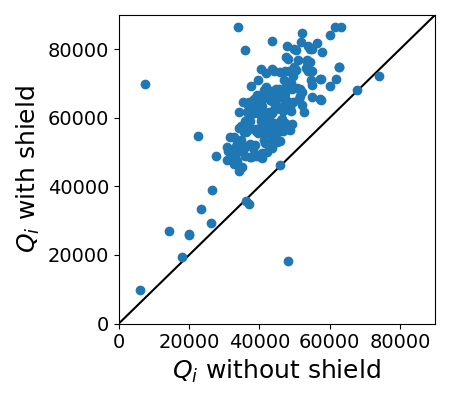}
    \includegraphics[width=0.49\textwidth]{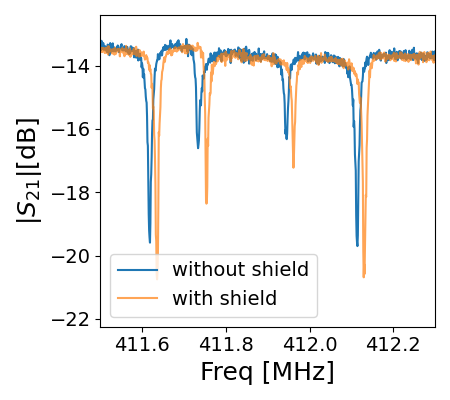}
%\vspace{-0.25in} 
    \caption{Left panel shows $Q_i$ for one feedline without magnetic shield compared to $Q_i$ with magnetic shield. Right panel shows example resonators for the two measurements. Using this magnetic shield with a shielding factor of $\sim$20 leads to $\sim$1.4 times larger $Q_i$ with negligible resonator frequency shifts.}
\label{fig:Q_vs_shield}
\vspace{-0.18in} 
\end{figure}

We also show the effect of magnetic shielding on the resonators in Figure~\ref{fig:Q_vs_shield}, by comparing the data taken without an external magnetic shield with the data taken after placing the magnetic shield outside the 300 K shell of the cryostat and thermal cycling through the detector critical temperature. The cylindrical magnetic shield is 1.5 mm thick Mu-metal, 889 mm tall, 447 mm diameter with an opening at the top, and is placed such that the array sits centered inside the cylinder $\sim$520 mm below the opening (right panel of Figure~\ref{fig:led_shield}). The measured shielding factor for static fields is $\sim$20 at this height. We find that using the magnetic shield leads to $\sim$1.4 times larger $Q_i$, while $\Delta f_0/f_0$ is $4\pm1\times 10^{-5}$, indicating the importance of magnetic shielding for the quality factors of these MKIDs. Ambient magnetic fields can get trapped in the MKID elements, suppressing local superconducting transitions in thin films, thereby increasing the number of quasiparticles and decreasing quality factors \cite{flanigan/2016, kutsuma/2018}. Further magnetic shielding studies with a Helmholtz coil are planned for the near future. For Prime-Cam, we expect a nominal magnetic shielding factor of $\sim$100 at the arrays based on simulations \cite{vavagiakis/2018b, ali/2020}. 

%add Q formula from seth siegel's thesis, how S21 was taken with the ROACH-2, how fitting was done, 

%Q vs mag shield, Q vs temperature,  

\section{Conclusion}
The characterization of the first 280 GHz MKID array for Prime-Cam is ongoing. We found that any resonator frequency shift due to the re-assembly process of the detector array package, in particular the alignment of the feedhorn array with the chokes that are close to the capacitors, is negligible compared to the effect of the LED mapper trimming process. We also showed that magnetically shielding the resonators can improve the quality factors and plan to quantify this in the near future. The LED position-frequency mapping and trimming process for the first 280 GHz array will soon take place, followed by a campaign of optical characterization work. Two more 280 GHz MKID arrays will be fabricated at NIST to fill one instrument module \cite{gudmundsson/2021}, and the possibility of using aluminum for the superconductor, and silicon-platelet feedhorns is being explored. Broadband instrument modules with 350 GHz and 850 GHz polarization-sensitive MKID arrays, and spectrometer instrument modules using MKID arrays coupled to Fabry-Perot interferometers are funded and in development. Testing and characterization of the first light receiver Mod-Cam with one instrument module is ongoing \cite{duell/2020}. The Prime-Cam cryostat is being machined and delivery is expected early 2022. Lastly, FYST is being constructed for first light with Mod-Cam in 2023.

\begin{acknowledgements}
SKC acknowledges support from NSF award AST-2001866. MDN acknowledges support from NSF award AST-2117631. ZBH acknowledges support from a NASA Space Technology Graduate Research Opportunities Award. YL acknowledges support from KIC Postdoctoral Fellowship. %This work was supported by the U.S. National Science Foundation through award 1440226. The development of multichroic detectors and lenses was supported by NASA grants NNX13AE56G and NNX14AB58G. The work of KTC and BJK was supported by NASA Space Technology Research Fellowship awards.
\end{acknowledgements}

\bibliographystyle{unsrt85}

\bibliography{JLTP_LTD19_publist}

\begin{thebibliography}{10}

\bibitem{ccat/2021}
{CCAT-Prime collaboration}, et~al.
\newblock {\em arXiv e-prints}, page arXiv:2107.10364, (2021).

\bibitem{bicep/2021}
P.~A.~R. {Ade}, et~al.
\newblock {\em \prl}, 127(15):151301, (2021).

\bibitem{dutcher/2021}
D.~{Dutcher}, et~al.
\newblock {\em \prd}, 104(2):022003, (2021).

\bibitem{choi/2020b}
Steve~K. {Choi}, et~al.
\newblock {\em \jcap}, 2020(12):045, (2020).

\bibitem{SO/2019}
Peter {Ade}, et~al.
\newblock {\em \jcap}, 2019(2):056, (2019).

\bibitem{hill/2018}
Charles~A. {Hill}, et~al.
\newblock volume 10708 of {\em Society of Photo-Optical Instrumentation
  Engineers (SPIE) Conference Series}, page 1070842, (2018).

\bibitem{choi/2020a}
S.~K. {Choi}, et~al.
\newblock {\em Journal of Low Temperature Physics}, 199(3-4):1089--1097,
  (2020).

\bibitem{mccarrick/2021}
Heather {McCarrick}, et~al.
\newblock {\em \apj}, 922(1):38, (2021).

\bibitem{hubmayr/2015}
J.~{Hubmayr}, et~al.
\newblock {\em Applied Physics Letters}, 106(7):073505, (2015).

\bibitem{austermann/2018}
J.~E. {Austermann}, et~al.
\newblock {\em Journal of Low Temperature Physics}, 193(3-4):120--127, (2018).

\bibitem{duell/2020}
Cody~J. {Duell}, et~al.
\newblock volume 11453 of {\em Society of Photo-Optical Instrumentation
  Engineers (SPIE) Conference Series}, page 114531F, (2020).

\bibitem{gordon/2016}
Samuel {Gordon}, et~al.
\newblock {\em Journal of Astronomical Instrumentation}, 5(4):1641003, (2016).

\bibitem{liu/2017}
X.~{Liu}, et~al.
\newblock {\em arXiv e-prints}, page arXiv:1711.07914, (2017).

\bibitem{flanigan/2016}
D.~{Flanigan}, et~al.
\newblock {\em Applied Physics Letters}, 109(14):143503, (2016).

\bibitem{kutsuma/2018}
Hiroki {Kutsuma}, et~al.
\newblock {\em Journal of Low Temperature Physics}, 193(3-4):203--208, (2018).

\bibitem{vavagiakis/2018b}
E.~M. {Vavagiakis}, et~al.
\newblock {\em Journal of Low Temperature Physics}, 193(3-4):288--297, (2018).

\bibitem{ali/2020}
Aamir~M. {Ali}, et~al.
\newblock {\em Journal of Low Temperature Physics}, 200(5-6):461--471, (2020).

\bibitem{gudmundsson/2021}
Jon~E. {Gudmundsson}, et~al.
\newblock {\em \ao}, 60(4):823, (2021).

\end{thebibliography}

\end{document}